# A GRID solution for Gravitational Waves Signal Analysis from Coalescing Binaries: preliminary algorithms and tests

F. Acernese[1,2], F. Barone[2,3], R. De Rosa[1,2], R. Esposito[2], P. Mastroserio[2], L. Milano[1,2], S. Pardi[1], K.Qipiani[2], G. Spadaccini[1,2]

[1]Università di Napoli "Federico II", Dipartimento di Scienze Fisiche, 80126 Napoli, Italy

[2]INFN – Sezione di Napoli, 80126 Napoli, Italy

[3]Università di Salerno, Dipartimento di Scienze Farmaceutiche, 84084 Salerno, Italy

The analysis of data coming from interferometric antennas for gravitational waves detection may require a huge amount of computing power. The usual approach to the detection strategy is to set-up computer farms able to perform several tasks in parallel, exchanging data through network links. In this paper a new computation strategy is presented. This strategy is based on the GRID environment that allows several geographically distributed computing resources to exchange data and programs in secure way, using standard infrastructures. The computing resources can be geographically distributed also on a large scale. Some preliminary tests were performed using a subnetwork of the GRID infrastructure, producing good results in terms of distribution efficiency and time duration.

## 1. INTRODUCTION

The detection of gravitational waves is certainly one of the most interesting fields of the modern physics: it will provide a strong proof of the general relativity theory, opening in this way a completely new channel of information on the dynamics and evolution of astrophysical objects [1]. In this scenario, the large scale terrestrial interferometric detectors like VIRGO [2], LIGO [3], GEO [4] and TAMA [5], will have a prominent role. In fact, these detectors will operate with large detection bands, typically spanning from 10 up to 10000 Hz, with a sensitivity of about $10^{-21}$ h/(Hz)$^{1/2}$ at 100 Hz, where h is the gravitational strain. In addition the very long baseline space interferometer LISA [6] will explore the detection frequency band from $10^{-5}$ up to 1 Hz.

For all these detectors, but especially for earth based antennas, the main problem to be solved in data analysis is the expected low signal to noise ratio. The low value of this quantity is due to the intrinsic weakness of gravitational signals with respect to the instrumental noise. To overcome this problem many efforts are being done in the development of suitable data analysis techniques. When the expected signals shape is known, the most promising technique seems to be the matched filtering, i.e. the correlation of the detector output with a set of theoretical waveform templates.

In the case of the VIRGO antenna the required computing power in detection of gravitational waves generated by a coalescing binary system of star is about 300 Gflops for masses ranging from 1.4 to 50 solar masses and with a signal to noise recovery of 90% [7][8].

These performances can be obtained using computer farms composed by several nodes connected each other through the network. This technical solution represents the only possible way for the on-line data analysis, when the data produced by the detector must be analysed in time. Furthermore a more precise analysis can be done off-line. In this case there is no time constraint and therefore the number of templates used can be increased and other parameters can be also included in the model. For this reason the off-line phase usually requires a very large amount of computing power that can be obtained only by adding more computing resources. The idea described in this paper is that of using the existing GRID environment as a platform to connect several geographically distributed computing and storing resources [9]. In this way the tasks needed for the off-line data analysis can be performed by dividing them in subtasks that will be executed on remote computers. The GRID structure also provides all the tools needed to collect back the results of the analysis.

## 2. THEORETICAL FRAMEWORK

### 2.1. Gravitational waves from coalescing binary systems

Coalescing binaries are systems of two close stars which orbits gradually decay from energy and angular momentum loss to gravitational waves. Usually a system of two Black Holes, two Neutron Stars or one Black Hole and one Neutron Star are considered. The coalescence can be divided into three phases: inspiral, merger and ringdown. During each phase a characteristic gravitational waveform is emitted.

The inspiral and ringdown waveforms are theoretically well known, while for the merger there is only a partial knowledge. The expressions of the two polarizations of the gravitational strain are:

$$h_{+,x} = \frac{2GM\eta}{c^2 r} \left(\frac{GM\omega}{c^3}\right)^{(2/3)}$$
$$\left[H_{+,x}^{(0)} + x^{1/2} H_{+,x}^{(1)} + x H_{+,x}^{(2)} + x^{3/2} H_{+,x}^{(3)} + x^2 H_{+,x}^{(4)} + \cdots\right]$$

where

$$M = m_1 + m_2 \, , \ \eta = \frac{m_1 m_2}{M^2} \, , \ x = \left(\frac{GM\omega}{c^3}\right)^{(2/3)}$$

**TULP001**

The $H_{+,x}^{(n)}$ are the terms of the n-order post newtonian expansion. Their expression is non trivial and can be found elsewhere [10]. They depend on the orbital plane position and on the actual orbital phase.

## 2.2. Matched filters for gravitational waves detection

Despite of their source, the detection of gravitational waves consists in recognizing a signal, h(t), in the noise background, n(t). The total signal, at any time t is s(t)=h(t)+n(t). Usually the waveform h(t) is theoretically known, and it is a function of parameters like the masses of the stars, the distance and the orientation of the orbital plane.

From a theoretical point of view, the best algorithm for signal extraction from noise, is the matched filtering [11]. The signal to noise ratio reaches its maximum value when this filter is applied to the data. Let us consider a filter K(t) to apply on data s(t).

$$Y = \int K(t)s(t)dt$$

The signal to noise ratio (SNR) is defined as:

$$\frac{S}{N} = \frac{\langle Y \rangle}{\sqrt{\langle Y_{s(t)=0} \rangle}}$$

If the shape of the signal h(t) is known, then the best filter (matched filter) is:

$$\widetilde{K}(f) = \widetilde{h} / S_h(f)$$

where $S_h(f)$ is the detector noise spectrum.

In principle one should compute all the SNR for each possible template with the data s(t). The largest SNR is the one corresponding to the template that is present in the signal. In practice a grid of templates with a minimal match as large as possible is generated, according to the available computing power.

For Coalescing Binaries the detection strategy consists in building a grid of templates based on the theoretical knowledge of the inspiral phase. Usually the templates are calculated for a couple of non spinning stars, using a restricted post newtonian approximation, discarding all multipolar components, but the quadrupole one.

In this approximation the frequency behavior of the gravitational waves is given by [10]:

$$h(f,\Phi_0,t_0,m_1,m_2) = f^{-7/6} e^{i\left[-\frac{\pi}{4}\Phi_0 + 2\pi t_0 + \Psi(f,m_1,m_2)\right]}$$

where $m_1$ and $m_2$ are the masses of the stars, $t_0$ and $\Psi_0$ are the time and the phase of the signal when it enters in the measurement band of the detector (usually the seismic cut-off, that for VIRGO is about 10 Hz), and $\Psi$ is the phase evolution of the binary system. Other parameters, not included in the templates, are the distance and the orientation of the orbital plan respect with respect to the detector. These parameters can be obtained after the filtering. In principle one should build a grid of templates with four parameters, but using the properties of Fourier Transform it is possible to take into account only the masses.

For the template placement the usual approach is the large match limit approximation. The match between two close templates a(t;p) and b(t;q), is given by:

$$M = \langle a(p) | b(q) \rangle = 4\Re\left[\int_0^\infty \frac{\widetilde{a}^*(f)\widetilde{b}(f)}{S_h(f)} df\right]$$

where p and q are a set of parameters of the template (masses or coalescing times). If the templates a and b are close it is possible to write:

$$q = p + \Delta p$$

and one can expand the match in series of Δp. If we stop the expansion to the second order and maximize the expression of matching, we can express the match as:

$$MM = 1 - g_{ij}\Delta p_i \Delta p_j$$

where the matrix $g_{ij}$ represents the metric of the template space. From the previous result, fixing MM, that is the minimal match between two close templates, it is possible to found the template spacing Δp.

The algorithm requires the calculation of the Fast Fourier Transform (FFT) of the data, the FFT of the templates, the computation of their correlation in frequency domain with respect to the instrumental noise, and the calculation of the Inverse FFT. These steps must be repeated for each template of the grid. Using the standard VIRGO conditions, for a total mass system from 0.25 to about 50 solar masses, with a SNR recovery of 90%, the number of templates is $9.8 \cdot 10^5$ requiring, for the real time process, about 300 Gflops: it is clear that parallel computing is needed to perform an on-line analysis.

The main expected problem for this kind of analysis, apart the large computational power, is the assumption that the templates are a good representation of the real phenomena. This is true if the system is far from the merging, the orbits are circular and the components do not spin.

If one of the previous conditions is not fulfilled, then the templates can give only an approximate description of the expected waveform. As a consequence we loose one of the main advantages of this optimal algorithm. On the other hand, if we take into account other relevant physical quantities, like eccentricity and spin, the number of templates increases exponentially and no on-line

**TULP001**

computation is possible, according to the nowadays status of the technology.

## 3. THE GRID INFRASTRUCTURE

A possible solution for getting high computing power needed for matched filter algorithm is that of adding independent computing resources, connected by some physical link in order to exchange data, procedures and results. The today technology gives the possibility of connecting several computers through the standard network at very low cost. In addition many packages are already available to program the algorithms required for the analysis in a parallel environment. In this way, by simply setting-up a farm with a sufficient number of nodes, it becomes possible to perform the analysis. The situation changes in the case of a very accurate analysis, when there is no special time constraint and the number of templates increases due to the introduction of other physical parameters that characterize the astrophysical system. In this case the computing power can raise beyond the capability of the farm used for the on-line analysis. Nevertheless, it is again possible to use resources connected through the network, but in this case it may often be more convenient to build up a network of geographically distributed resources, that can be used by the local users as well as by other groups as the computing power becomes available.

The GRID structure represents a suitable solution for the distributed computing. It allows to share and coordinate the use of resources within large, dynamic and multi-institutional communities [9].

In this environment the DataGRID project is a European Community project to develop a GRID testbed infrastructure on European scale [12]. The typical operations performed inside the GRID are associated to a physical element of it:

- the execution of programs is performed by dedicated computer called Computing Element (CE); more precisely each CE can manage several computer, called Worker Nodes (WN), where the programs run physically;

- the access to data is ensured by a class of computer, usually equipped with a large amount of mass storage, called Storage Element (SE);

- the distribution of data between the computing resources is ensured by the network;

- the user can submit his jobs through a computer called User Interface (UI).

All the infrastructure does not appear to the final user who can submit a job without specifying where it has to run and where the data to be processed are stored. In addition, the data transfer through the network is secure, due to a very effective authentication mechanism.

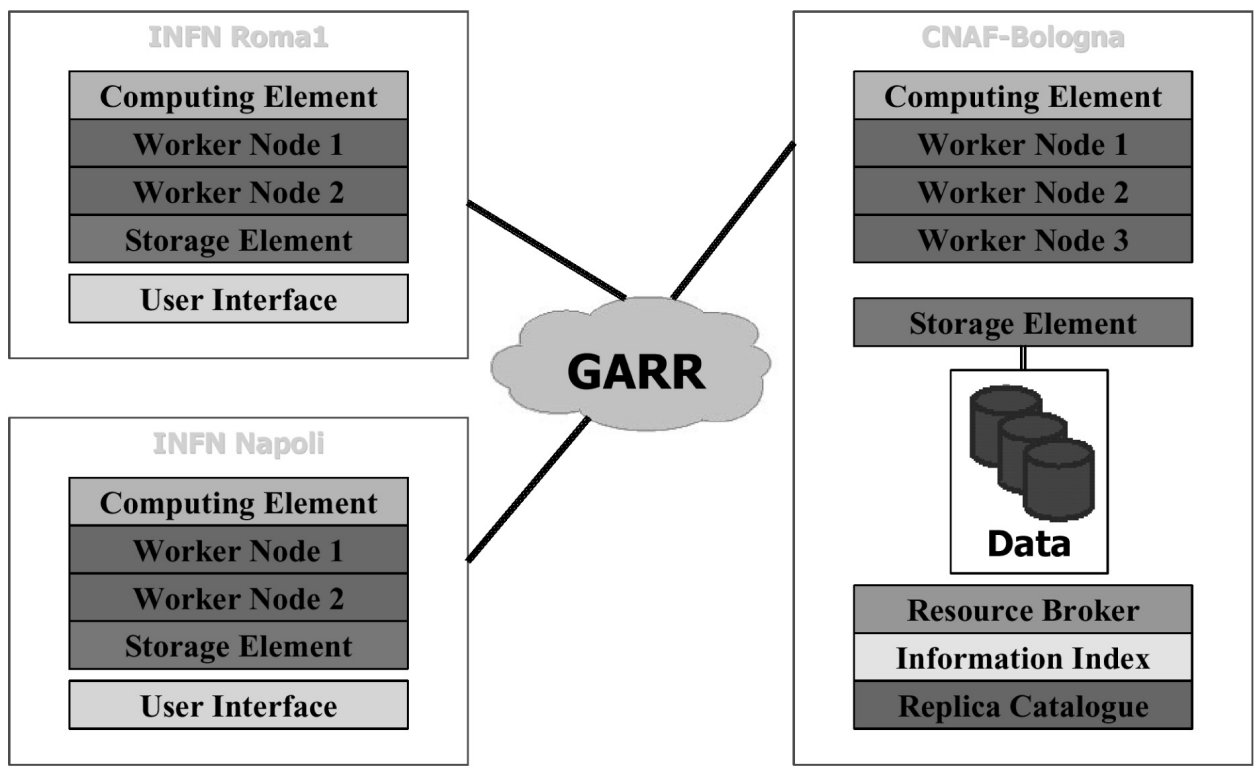

Figure 1: Scheme of the Virgo GRID structure

**TULP001**

## 4. GRID TEST FOR MATCHED FILTERS

A test of geographically distributed computation of matched filters for coalescing binaries were performed within the INFN GRID infrastructure, using a subnetwork composed by computing resources located in CNAF - Bologna, INFN - Napoli and INFN - Roma1.

The resources were subdivided in the following way:

- Bologna: 1 CE, 3 WN, 1 SE;
- Napoli: 1 CE, 2 WN, 1 SE, 1 UI;
- Roma: 1 CE, 2 WN, 1 SE, 1 UI.

This configuration allows the submission of independent jobs from Napoli and Rome.

A picture of the Virgo GRID configurations is reported in figure 1.

The conditions used to test the procedure are the following:

- Algorithm: standard matched filters
- Templates generated at PN order 2 with Taylor approximants

The data used to test the procedure have the following characteristics:

- Data simulated at 20 kHz
- Each data frame is 1 second long
- Total data length: 600 s

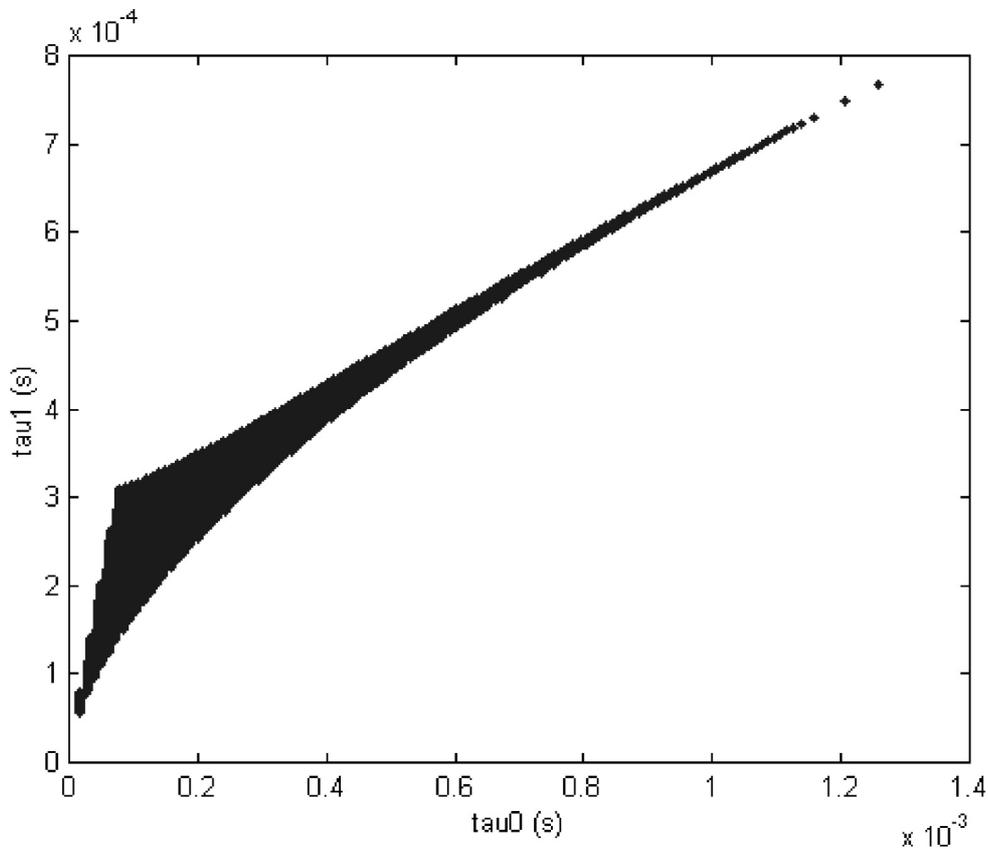

Figure 2: Map of the templates used for the test

The data analysis condition were the following:

- raw data resampled at 2 kHz
- lower frequency: 60 Hz
- upper frequency: 1 kHz
- search space: 2 – 10 solar masses
- minimal match: 0.97
- Number of templates: ~ 40000

A graphic of the templates parameters space is shown in Fig.2. We choose, as parameters, the coalescing times $\tau_0$ and $\tau_1$.

The analysis procedure was divided in two steps:

- Step 1

The data were extracted from CNAF-Bologna Mass Storage System. The extraction process reads the VIRGO standard frame format, performs a simple resampling and publishes the selected data file on the Storage Element;

- Step 2

The search was performed dividing the template space in 200 subspace and submitting from Napoli User Interface a job for each template subspace.

Each job reads the selected data file from the Storage Element (located at CNAF-Bologna) and runs on the Worker Nodes selected by Resource Broker in the VIRGO VO.

Finally, the output data of each job were retrieved from Napoli User Interface

## 5. CONCLUSIONS

We have successfully verified that multiple jobs can be submitted and the output retrieved with a small overhead time. This demonstrates that a geographically distributed computing can be very effective for heavy data analysis algorithms, in particular for the coalescing binaries off-line data analysis. The GRID infrastructure does not seem to affect in a significant way the full computation time. On the contrary it provides a convenient environment for the development of extended computational projects. Of course other tests are needed to found the optimal distribution strategy for both data and algorithms.

### References

[1] C. W. Misner, K. S. Thorne, J. A. Wheeler, Gravitation (Freeman & Co., San Francisco, 1973).

[2] C. Bradaschia et al., "The VIRGO Project, Final Design of the Italian-French large base interferometric antenna of gravitational wave detection", Proposal to INFN Italy and CNRS France, 1989, 1992, 1995.

[3] R.E. Vogt, R.W. Drever, F.J. Raab, K.S. Thorne, "Proposal for the construction of a large interferometric detector of gravitational waves", Proposal to the National Science Foundation, California Institute of Technology, Pasadena, California, USA, 1989.

[4] Hough et al., "Proposal for a joint german-british interferometric gravitational wave detector", MPQ 147, Max Planck Institut für Quantenoptik, Munich, Germany, 1989

[5] unpublished, see http://tamago.mtk.nao.ac.jp (1996)

[6] P. Bender, et al. "LISA: Laser Interferometer Space Antenna for the detection and the observation of gravitational waves},MPQ 208, Max Planck Institut für Quantenoptik, Munich, Germany, 1996.

[7] B. Owen, 1996, Phys. Rev. D, 53, 6749.

[8] P. Canitrot, L. Milano, A, Vicere', VIR-NOT-PIS-1390-149, 2000

[9] I. Foster, 2002, Physics Today, 55, 42

[10] L. Blanchet, T. Damour, B. R. Iyer, 1995, Phys Rev. D, 51, 5360

[11] C. W. Helstrom, Statistical Theory of Signal Detection, 2nd ed. (Pergamon Press, London, England, 1968)

[12] Chervenak, I. Foster, C. Kesselman, C. Salisbury, S. Tuecke, 2001, Journal of Network and Computer Applications, 23, 187